\documentclass[aps,prb,twocolumn,floatfix,citeautoscript,hyperref=pdftex]{revtex4-1}
\usepackage{times}
\usepackage{graphicx}
\usepackage{dcolumn}
\usepackage{bm}
\usepackage{color}
\usepackage{amsmath}
\usepackage{amssymb}
\newcommand {\beq} {\begin{equation}}
\newcommand {\eeq} {\end{equation}}
\newcommand {\bqa} {\begin{eqnarray}}
\newcommand {\eqa} {\end{eqnarray}}

\usepackage[colorlinks=true]{hyperref}

\begin{document}

\title{Interplay of finite-energy and finite-momentum superconducting pairing}

\author{Debmalya Chakraborty}
\affiliation{Department of Physics and Astronomy, Uppsala University, Box 516, S-751 20 Uppsala, Sweden}

\author{Annica M. Black-Schaffer}
\affiliation{Department of Physics and Astronomy, Uppsala University, Box 516, S-751 20 Uppsala, Sweden}

\begin{abstract}

Understanding the nature of Cooper pairs is essential to describe the properties of superconductors. The original proposal of Bardeen, Cooper, and Schrieffer (BCS) was based on electrons pairing with same energy and zero center-of-mass momentum. With the advent of new superconductors, different forms of pairing have been discussed. In particular, Cooper pairs with finite center-of-mass momentum have received large interest. Along with such finite-momentum pairs, pairing of electrons at different energies is also central to understanding some superconductors. Here, we investigate the interplay of finite-momentum and finite-energy Cooper pairs considering two different systems: a conventional $s$-wave superconductor under applied magnetic field and a $d$-wave finite-momentum pairing state in the absence of magnetic field relevant to correlated superconductors. Investigating both these systems, we find finite-energy pairs persisting independently of finite-momentum pairing, and that they lead to odd-frequency superconducting correlations. We contrast this finding by showing that the even-frequency correlations are predominantly driven by zero-energy pairs for most frequencies. We further calculate the Meissner effect and find that odd-frequency correlations are essential for correctly describing the Meissner effect.

\end{abstract}

\maketitle

\section{Introduction}\label{sec:Intro}

\begin{figure}[t]
\includegraphics[width=1.0\linewidth]{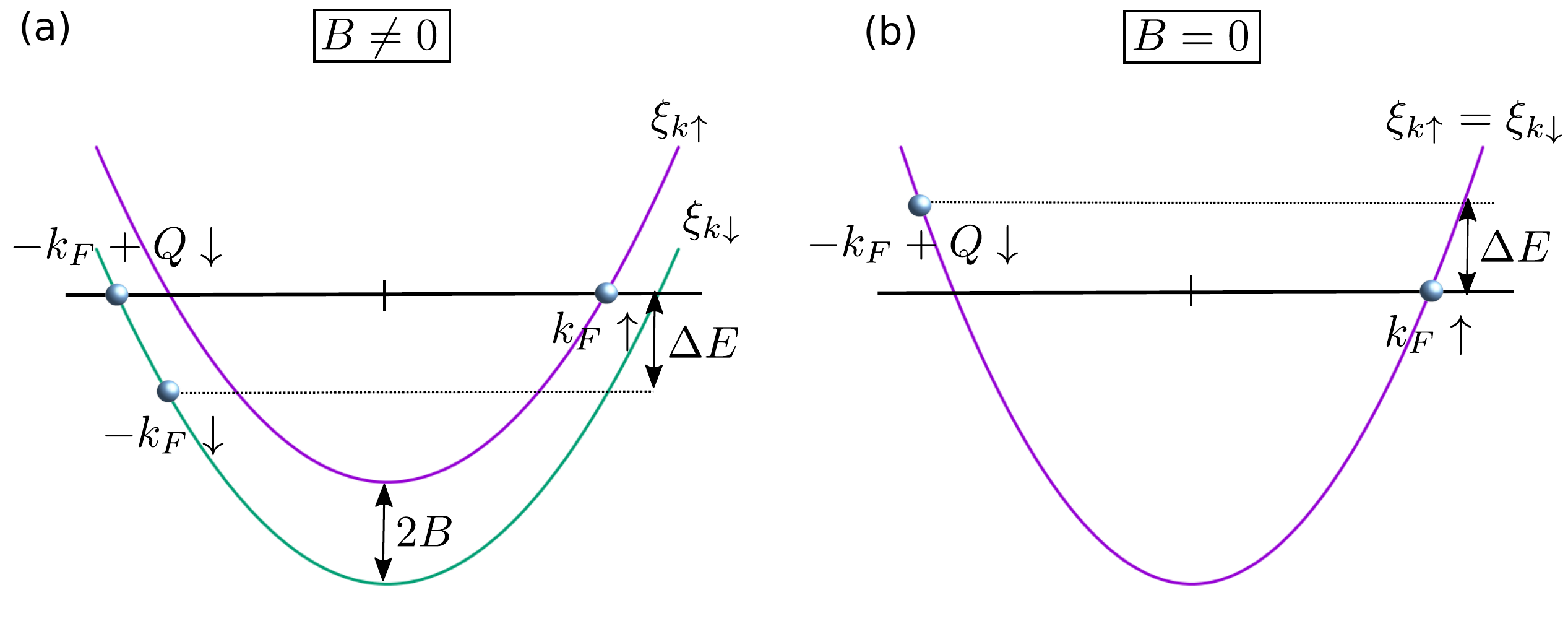} \caption{Schematic showing finite-energy and finite-momentum Cooper pairs in the presence of magnetic field $B\ne 0$ (a) and in the absence of magnetic field $B=0$ (b). A typical 1D band dispersion for $\uparrow$-spin, $\xi_{k\uparrow}$, and $\downarrow$-spin, $\xi_{k\downarrow}$, electrons are shown. Black horizontal line shows the Fermi level with $k_{F}$ being the Fermi momentum. $Q$ denotes the total center-of-mass momentum and $\Delta E$ denotes the finite-energy of the Cooper pair.}
\label{fig:schemfig} 
\end{figure}

Many superconductors can be described by the famous Bardeen, Cooper, and Schrieffer (BCS) theory \cite{Bardeen57}. The BCS theory in its original form is based on the formation of Cooper pairs of electrons near the Fermi level with opposite momentum $k$ and $-k$ and opposite spins $\uparrow$ and $\downarrow$ in a singlet configuration. In the presence of time-reversal symmetry, electrons with opposite momentum and spins have equal energy in the band dispersion, leading to pairing between electrons with the same energy. However,  BCS theory is also applicable in the absence of time-reversal symmetry, e.g.~in the presence of an applied magnetic field. In this case, the electrons forming the Cooper pairs are at different energies in the band dispersion, see Fig.\ref{fig:schemfig}(a). Such pairs of electrons can aptly be called finite-energy Cooper pairs. Finite-energy pairing has also recently been proposed to intrinsically occur in monolayer transition-metal dichalcogenides \cite{Tang21} and $j=3/2$ superconductors \cite{Bahari22}. Another known variant of Cooper pairs is when electrons pair with momentum different from opposite momenta. In this scenario, the total center-of-mass momentum of the Cooper pairs is non-zero and are hence commonly known as finite-momentum pairs. Finite-momentum pairs have recently attracted renewed attention due to their possible emergence in several intensively studied superconductors such as cuprates \cite{Hamidian16,Edkins19}, transition-metal dichalcogenides \cite{Liu21}, iron-based superconductors \cite{Kasahara20}, and kagome metals \cite{Chen21}. Understanding the interplay between both finite-energy and finite-momentum Cooper pairs will likely both provide deeper insights into the pairing symmetry of existing superconductors and open pathways to discover future superconductors with exotic properties.

A simple example where both finite-energy and finite-momentum Cooper pairs are present is a conventional spin-singlet $s$-wave superconductor under an applied magnetic field. The Zeeman effect of the magnetic field splits the Fermi surface of the $\uparrow$- and $\downarrow$-spin electrons, thereby straining the singlet configuration of the Cooper pairs. However, the condensation energy gain due to the formation of Cooper pairs still enables superconductivity to survive weak magnetic fields. In this weak field regime, BCS spin-singlet pairs ($k_F \uparrow,-k_F \downarrow$) are formed by electrons at different energies, thus turning into finite-energy pairs, see Fig.\ref{fig:schemfig}(a), where $k_F$ is the Fermi momentum. For higher magnetic fields, an interplay between the magnetization energy and the condensation energy is known to result in the formation of finite-momentum Cooper pairs ($k_F \uparrow,-k_F+Q \downarrow$). This phenomenon was originally illustrated independently by Fulde-Ferrell (FF) \cite{Fulde64} and Larkin-Ovchinnikov (LO) \cite{Larkin64}. While FF proposed finite-momentum Cooper pairs with $Q$-momentum modulations in the phase factor of the superconducting (SC) order parameter, LO suggested finite-momentum Cooper pairs with modulations in the amplitude of the SC order parameter. Notably, in the finite-momentum SC state ($k_F \uparrow,-k_F+Q \downarrow$) electrons are no longer at different energies, as illustrated in Fig.\ref{fig:schemfig}(a). This may give us the naive expectation that finite-momentum pairs are always zero-energy pairs, such that the finite-momentum pairing relievies the superconductor of finite-energy pairs. However, the schematic drawn in Fig.\ref{fig:schemfig}(a) represents a one-dimensional (1D) band dispersion. If we instead consider a system in 2D or 3D, $k_F$ lie either on a line or a surface, thus giving the possibility of the ($k_F \uparrow,-k_F+Q \downarrow$) electrons not being at same energies. This brings us to the first concrete question we aim to answer in this work: Do finite-energy Cooper pairs generally exist in a finite-momentum SC state?

A finite-momentum SC state has  recently been  found experimentally in correlated electron systems \cite{Hamidian16,Liu21,Chen21}, even in the absence of applied magnetic field, often referred to as a pair density wave state. This state is theoretically proposed to be spontaneously formed due to the effects of electronic correlations \cite{Agterberg20,Wang15b,Chakraborty19,Waardh17,Choubey20}. Such a state with ($k_F \uparrow,-k_F+Q \downarrow$) pairs actually necessarily have finite-energy Cooper pairs, since the $\uparrow$ and $\downarrow$ electrons are degenerate, for an illustration see Fig.\ref{fig:schemfig}(b). This leads to the second question we aim to address in this work: What is the nature of the finite-energy pairs generated due to the presence of finite-momentum pairs in the absence of applied magnetic field?

Till now we have only discussed Cooper pairs with unequal energy or unequal momentum, but still formed at equal time. A different form of SC pairing can exist where two electrons pair at unequal times. Such unequal time Cooper pairs gives the possibility that the pair wavefunction becomes odd under the exchange of the electron time coordinates or, equivalently, odd in frequency \cite{Berezinskii74, Kirkpatrick91, Balatsky92, Schrieffer94, Bergeret05, Linder19}. Odd-frequency pairing has been instrumental in understanding several non-intuitive experimental findings in superconductor-ferromagnet heterostructures \cite{Bergeret01,Buzdin05,Eschrig08,DiBernardo15,Krieger20} and is also proposed to exist in several bulk superconductors \cite{Triola18,Chakraborty21,Dutta21,Cayao21}, particularly in multi-band systems \cite{Black-Schaffer13,Komendova15,Asano15,Komendova17,Triola20,Schmidt20}. With time, frequency, and energy being closely related in quantum systems, this begs a third question we aim to address in this work: Do finite-energy, and also finite-momentum, superconductors host odd-frequency SC correlations?

Notably, the odd-frequency pairing discussed in the recent literature mostly involves odd-frequency SC correlations, which are distinct from the odd-frequency order parameter originally proposed \cite{Berezinskii74, Kirkpatrick91, Balatsky92, Schrieffer94} in the context of odd-frequency superconductivity. But, such intrinsic odd-frequency superconductivity may be thermodynamically unstable, \cite{Heid95} due to a most often found paramagnetic, or negative, Meissner response \cite{Hashimoto01,Bergeret01,Bernardo15}. However, a diamagnetic Meissner effect has been shown to be restored if odd-frequency pairs also have finite-momentum \cite{Hoshino14,Hoshino16}. This raises the fourth and last question we aim to address in this work: How do odd-frequency SC correlations affect the Meissner response in a superconductor with finite-energy, or in combination with finite-momentum, pairs?

To answer all these questions, we consider in this work  two different systems. The first system is a conventional spin-singlet $s$-wave superconductor in the presence of applied magnetic field, where the magnetic field eventually give rise to finite-momentum pairs.
By self-consistently solving the resulting Hamiltonian, we answer the first question by showing that finite-energy Cooper pairs are the only possibility in the BCS phase, but also and clearly prevalent in the finite-momentum FF phase. Moreover, we find that odd-frequency SC correlations exist in both BCS and FF phases and, notably, are only generated due to the presence of finite-energy pairs. In contrast, we find that  even-frequency correlations are mainly dominated by (near) zero-energy pairs for most frequencies. This shows both that odd-frequency SC correlations are intricately linked to finite-energy pairing and thus answers our third question. 
We also show the importance of these odd-frequency SC correlations in the Meissner effect by calculating the superfluid weight. While we find that odd-frequency correlations give a negative contribution to the superfluid weight, their inclusion is essential to correctly describe the magnetic field evolution of the superfluid weight and that in total the Meissner effect is still diamagnetic. This provides a clear answer to our fourth question. 
Finally, to generalize our results, we also study finite-momentum pairing in the absence of magnetic fields. We do this by studying an unconventional $d$-wave superconductor where we find a finite-momentum FF $d$-wave state spontaneously formed by self-consistently solving a pair hopping model proposed in the context of cuprates \cite{Waardh17}. Also in this finite-momentum system, we find coexisting finite-energy pairing and odd-frequency correlations that are directly related to these finite-energy pairs. This both answers our second question and, importantly, generalizes our other results and conclusions derived from the conventional superconductor in a magnetic field. 

We organize the rest of the article in the following way. In Sec.~\ref{sec:Zeeman} we discuss the case of a conventional superconductor in the presence of an applied magnetic field. We first give the details of the model Hamiltonian and discuss the procedure of self-consistency to obtain the ground state in Sec.~\ref{sec:modelZeeman}. We then find the SC correlations in the ground state in Sec.~\ref{sec:pctf} and then relate the obtained correlations to finite-energy and finite-momentum Cooper pairs in Sec.~\ref{sec:zeemanresults}. We then investigate the case of a spontaneously formed finite-momentum $d$-wave FF state in the absence of magnetic field in Sec.~\ref{sec:pairhopping}. Here we first give the details of the model and procedure of the self-consistency in Sec.~\ref{sec:modelpair} and then show the obtained SC correlations in Sec.~\ref{sec:correlationspair}. After that we discuss the effects of SC correlations in the Meissner effect in Sec.~\ref{sec:meissner}. Finally, we summarize our findings and also discuss the possible interplay of finite-energy and finite-momentum pairing in other systems in Sec.~\ref{sec:Discussion}.
 
\section{Finite-energy pairing in the presence of magnetic field}\label{sec:Zeeman}

\subsection{Model and ground state}\label{sec:modelZeeman}

In the anticipation that finite-momentum superconductivity eventually appears in the presence of magnetic field, we start with a generic mean-field Hamiltonian in 2D allowing for the possibility of forming finite-momentum SC pairs:
\begin{eqnarray}
H_{B}&=&\sum_{k,\sigma} \left(\xi_{k}+\sigma B \right) c_{k \sigma}^{\dagger} c_{k \sigma} \nonumber \\
&+& \sum_{k} \left( \Delta^{Q}_{k} c_{-k+Q/2 \downarrow} c_{k+Q/2 \uparrow} + \textrm{H.c.} \right)+\text{constant}.
\label{eq:Hamil}
\end{eqnarray}
Here $c_{k \sigma}^{\dagger}$ ($c_{k \sigma}$) is the creation (annihilation) operator of an electron with spin $\sigma$ and momentum $k$, $\xi_{k}$ is the electron band dispersion, $B$ is the applied magnetic field causing a Zeeman splitting of the electron energies with the magnetic moment of the electron $\mu_0$ taken to be unity, $\Delta^{Q}_k$ is the spin-singlet $s$-wave SC order parameter, and $Q$ is the total center-of-mass momentum of the Cooper pairs. For simplicity, we set the band dispersion $\xi_{k}=-2t(\cos(k_x)+\cos(k_y))-\mu$, where $t=1$ is the energy unit, and tune $\mu$ such that the average density of electrons $\rho=\sum_{k,\sigma}\langle c^{\dagger}_{k\sigma}c_{k\sigma}\rangle$ is kept fixed to a general value of $0.7$. $\Delta^{Q}_k$ is obtained by the self-consistency relation,
\begin{equation}
\Delta^{Q}_k=\sum_{k^{\prime}}V_{k,k^{\prime}} \langle c_{k^{\prime}+Q/2 \uparrow}^{\dagger} c_{-k^{\prime}+Q/2 \downarrow}^{\dagger} \rangle,
\label{eq:sc}
\end{equation}
where $V_{k,k^{\prime}}$ is the interaction strength driving the SC order. In this section we only consider $s$-wave superconductors with $\Delta^{Q}_k=\Delta^{Q}_0$, and hence we set $V_{k,k^{\prime}}=-V$, a constant independent of momentum. We achieve the BCS zero-momentum superconductivity if $Q=0$ and finite-momentum superconductivity if $Q\ne 0$ \cite{Cui06}. We take $V=2.5$ for obtaining a large SC gap to make the analysis clear. We work with a square lattice of size $N=1000\times 1000$, a value large enough to mimic the thermodynamic limit and capturing the relevant values of $Q$.

The Hamiltonian in Eq.~\eqref{eq:Hamil} can be written in a matrix form using the basis $\Psi^{\dagger}=\left(c_{k+Q/2 \uparrow}^{\dagger},c_{-k+Q/2 \downarrow}\right)$ as,
\begin{equation}
H_{B}=\sum_{k} \Psi^{\dagger} \hat{H}_{B} \Psi+\text{constant},
\end{equation}
with
\begin{equation}
\hat{H}_{B}=\left(\begin{array}{cc} \xi_{k+Q/2\uparrow} & \Delta^{Q}_{0} \\
\Delta^{Q}_{0} & -\xi_{-k+Q/2\downarrow} \\
\end{array}\right),
\label{eq:Hamilmat}
\end{equation} 
where now $\xi_{k \sigma}=\xi_{k}+\sigma B$ and $\Delta^{Q}_{0}$ is taken to be real-valued without any loss of generality.
We diagonalize the Hamiltonian $\hat{H}_{B}$ for a fixed $Q$ and solve for the self-consistency condition Eq.~\eqref{eq:sc} iteratively using the eigenvalues and the eigenvectors of Eq.~\eqref{eq:Hamilmat}. However, the self-consistent solutions of $\Delta^{Q}_{0}$ for a particular chosen $Q$ does not guarantee a global energy minimum. The global minimum, and thus the ground state solution $\Delta^{Q}_{0}$, can only be obtained by calculating the ground state energy $E=\sum_{k,\sigma}\xi_{k\sigma}\langle c^{\dagger}_{k\sigma}c_{k\sigma} \rangle-(\Delta^{Q}_0)^2/V+\mu\rho$ as a function of $Q$, and finding the optimal $Q$ that minimizes $E$. We note that $Q$ is a vector with two possible directions in 2D. Here we consider only uniaxial $Q$ along the $x$-axis and call it $Q_x=Q$ for notation simplicity. Since we consider only $s$-wave superconductivity in this section, other directions of $Q$ are expected to give similar results. 

We perform the above procedure of finding the ground state solution $\Delta^{Q}_{0}$ for different values of magnetic field to find the phase diagram as a function of $B$. For $B<B_{c1}\approx 0.35$, we find the $Q=0$ solution to be the ground state, showing the stability of the BCS phase in this range of $B$. In the range $B_{c1}<B<B_{c2}\approx 0.58$, we find that $Q\ne 0$ instead gives the global minimum in energy. Thus, in this intermediate range of magnetic fields, finite-momentum SC state is stable. Since we consider a single value of $Q$ in Eq.~\eqref{eq:Hamil}, the SC order in real space has modulations only in the phase factor and not in the amplitude, i.e.~it is an FF phase. To consider a LO phase with modulations in the amplitude of the SC order parameter, at least two $Q$ need to be considered. However, to avoid any complexity arising from emergent charge density wave orders in the LO phase, we do not consider the LO phase in this work, only focusing on the FF phase. Finally, for $B>B_{c2}$, there are no non-zero solutions of $\Delta^{Q}_{0}$, which means that the system is in normal, i.e.~non-superconducting, state. To summarize, $B_{c1}$ demarcates the transition of the BCS state to the FF state and $B_{c2}$ demarcates the transition of the FF state to the normal state.

\subsection{Superconducting correlations}\label{sec:correlationsZeeman}

\subsubsection{Theoretical framework and analytical results}\label{sec:pctf}

Having obtained the ground state of the Hamiltonian in Eq.~\eqref{eq:Hamil}, we next look at the SC pair correlations. The SC pair correlator is given by $F_{k,-k}(\tau)=-\langle T_{\tau}c_{k+Q/2 \uparrow}^{\dagger}(\tau) c_{-k+Q/2 \downarrow}^{\dagger}(0) \rangle$, where $\tau$ is the imaginary time and $T_{\tau}$ is the $\tau$-ordering operator. We here choose to not indicate the $Q$ dependence in $F_{k,-k}(\tau)$ for notational simplicity. After Fourier transforming, $F_{k,-k}(\tau)$ can be written as $F_{k,-k}(i\omega)$, where $\omega$ are fermionic Matsubara frequencies. The SC pair correlator $F_{k,-k}(i\omega)$ can be obtained directly from the off-diagonal part of the Green's function $G$, given by $G^{-1}(i\omega)=i\omega-\hat{H}_{B}$. Thus, by using the Hamiltonian in Eq.~\eqref{eq:Hamilmat}, the Green's function is obtained by inverting the $2\times2$ matrix $G^{-1}(i\omega)$ and the pair SC correlator is given by,
\begin{equation}
F_{k,-k}(i\omega)=G_{12}(i\omega)=F^{e}_{k,-k}(i\omega)+F^{o}_{k,-k}(i\omega),\label{eq:anomolous} \\
\end{equation}
where
\begin{eqnarray}
F^{e}_{k,-k}(i\omega)&=&\frac{-\Delta^{Q}_0\left( \xi_{k+Q/2\uparrow}\xi_{-k+Q/2\downarrow}+(\Delta^{Q}_0)^2+\omega^2 \right)}{D},\label{eq:feven}\\
F^{o}_{k,-k}(i\omega)&=&\frac{i\omega\Delta^{Q}_0\left( \xi_{k+Q/2\uparrow}-\xi_{-k+Q/2\downarrow} \right)}{D},\label{eq:fodd} \\
D&=&\left( \xi_{k+Q/2\uparrow}\xi_{-k+Q/2\downarrow}+(\Delta^{Q}_0)^2+\omega^2 \right)^2 \nonumber \\
&&+\omega^2\left( \xi_{k+Q/2\uparrow}-\xi_{-k+Q/2\downarrow} \right)^2. \label{eq:D}
\end{eqnarray}
We have here decomposed $F_{k,-k}(i\omega)$ into its even- ($F^{e}_{k,-k}(i\omega)$) and odd-frequency ($F^{o}_{k,-k}(i\omega)$) components, as clearly $F^{e}_{k,-k}(i\omega)$ and $F^{o}_{k,-k}(i\omega)$ have even and odd-frequency dependence, respectively, in the nominator, while the denominator $D$ is an even function of frequency.

Already at this stage, we can relate the SC correlations with finite-energy Cooper pairs. In the BCS phase, $(k,\uparrow)$ electrons pair with $(-k,\downarrow)$ electrons and $Q=0$. In the absence of magnetic field $B$ and for inversion symmetric superconductors, $\xi_{k\uparrow}=\xi_{-k\downarrow}$. From Eq.~\eqref{eq:fodd} we then find $F^{o}_{k,-k}(i\omega) \propto \xi_{k\uparrow}-\xi_{-k\downarrow}=0$. Hence, in the absence of magnetic field, there are no odd-frequency correlations, whereas $F^{e}_{k,-k}(i\omega)$ is still finite, as seen from Eq.~\eqref{eq:feven}. In the presence of magnetic field, but still in the BCS phase, $(k,\uparrow)$ electrons still pair with $(-k,\downarrow)$ electrons and $Q=0$, but now $\xi_{k\uparrow}\ne \xi_{-k\downarrow}$. As a result, it is electrons of different energies in the normal state that form the Cooper pairs. In this case, $F^{o}_{k,-k}(i\omega)\propto \xi_{k\uparrow}-\xi_{-k\downarrow}\ne 0$ and thus the odd-frequency correlations are proportional to the energy difference of the electrons pairing, or equivalently, to the existence of finite-energy pairs. Finally, in the FF state, $(k+Q/2,\uparrow)$ electrons pair with $(-k+Q/2,\downarrow)$ electrons with the difference in their energies being $\xi_{k+Q/2\uparrow}- \xi_{-k+Q/2\downarrow}$. Also in this case $F^{o}_{k,-k}(i\omega)\propto \xi_{k+Q/2\uparrow}- \xi_{-k+Q/2\downarrow}$, see Eq.~\eqref{eq:fodd}, and the odd-frequency correlations are still proportional to the energy difference of the electrons pairing. The above analysis thus shows that odd-frequency correlations necessarily require the formation of finite-energy pairs in the ground state, whereas even-frequency correlations exist even if only zero-energy pairs are present. Already here, we thus answer the third question posed in Sec.~\ref{sec:Intro} by establishing a direct analytical relation of odd-frequency SC correlations with finite-energy Cooper pairs. A remarkable consequence of this result is that the odd-frequency correlations act as a direct measure of finite-energy Cooper pairs as these correlations are directly proportional to the energy difference of the electrons forming the Cooper pairs. In fact, we will use this feature in Sec.~\ref{sec:zeemanresults} to answer the first question on whether finite-energy pairs are present in the finite-momentum state.

We next derive the spin properties of the pair correlation functions. Due to the fact that a pair correlation function should always satisfy the Fermi-Dirac statistics, the correlation function under a joint operation of spin permutation (S), momentum exchange or parity (P), and relative time permutation (T) of the individual electrons should satisfy $SPT=-1$. From Eqs.~\eqref{eq:feven} and \eqref{eq:fodd}, we find that under $P$ $F^{e}_{-k,k}(i\omega)=F^{e}_{k,-k}(i\omega)$, while $F^{o}_{-k,k}(i\omega)=-F^{o}_{k,-k}(i\omega)$. Next, spin-singlet correlations are always odd under $S$, while spin-triplet are even. So, in order to satisfy $SPT=-1$, the spin-singlet component of the even-frequency pair correlations $F^{e}_{s}(k,i\omega)$ can be obtained by taking an even combination of $F^{e}_{-k,k}(i\omega)$ and $F^{e}_{k,-k}(i\omega)$, as even-frequency spin-singlet correlations are required to be even under $P$. Likewise, the spin-triplet component $F^{e}_{t}(k,i\omega)$ can be obtained by taking the odd momentum combination. Following the same argument, the spin-singlet $F^{o}_{s}(k,i\omega)$ and triplet $F^{o}_{t}(k,i\omega)$ components of the odd-frequency pair correlations are obtained by considering the odd and even combinations of $F^{o}_{-k,k}(i\omega)$ and $F^{o}_{k,-k}(i\omega)$, respectively. Using the above analysis, we arrive at,
\begin{eqnarray}
&&F^{e}_{s}(k,i\omega)=\frac{F^{e}_{k,-k}(i\omega)+F^{e}_{-k,k}(i\omega)}{2}, \label{eq:evensing} \\
&&F^{e}_{t}(k,i\omega)=\frac{F^{e}_{k,-k}(i\omega)-F^{e}_{-k,k}(i\omega)}{2}, \label{eq:eventrip} \\
&&F^{o}_{s}(k,i\omega)=Im\left(\frac{F^{o}_{k,-k}(i\omega)-F^{o}_{-k,k}(i\omega)}{2}\right), \label{eq:oddsing} \\
&&F^{o}_{t}(k,i\omega)=Im\left(\frac{F^{o}_{k,-k}(i\omega)+F^{o}_{-k,k}(i\omega)}{2}\right), \label{eq:oddtrip}
\end{eqnarray}
where we have, for plotting purposes, taken the imaginary part in the last two lines since $F^{o}_{k,-k}(i\omega)$ is purely imaginary, see Eq.~\eqref{eq:fodd}. In the BCS phase where $Q=0$, using Eqs.~\eqref{eq:feven}-\eqref{eq:oddtrip}, we see that only spin-singlet even-frequency $F^{e}_{s}(k,i\omega)$ and spin-triplet odd-frequency $F^{o}_{t}(k,i\omega)$ correlations persist, whereas in the FF phase with finite $Q$, all the components in Eqs.~\eqref{eq:evensing}-\eqref{eq:oddtrip} are generally finite. Since our main focus is to compare the BCS and the FF phases, in the rest of this section, we primarily focus on $F^{e}_{s}(k,i\omega)$ and $F^{o}_{t}(k,i\omega)$, since they are both finite in both the BCS and FF phases. We also note that in the FF phase, the spin-singlet and -triplet components of the even-frequency correlations have very similar magnitudes and frequency dependence and the same is true for the spin-singlet and -triplet components of odd-frequency correlations. This gives additional good reason to only present our results for $F^{e}_{s}(k,i\omega)$ and $F^{o}_{t}(k,i\omega)$. We further characterize the total momentum contribution by defining the following two momentum sums,
\begin{equation}
F^{e/o}_{s/t}(i\omega)=\sum_{k} \left| F_{s/t}^{e/o}(k,i\omega) \right|, \label{eq:EO_modksum} 
\end{equation}
which quantifies the momentum-averaged absolute values and
\begin{equation}
{F_{*}}^{e/o}_{s/t}(i\omega)=\sum_{k} F_{s/t}^{e/o}(k,i\omega), \label{eq:EO_ksum}
\end{equation}
which quantifies the momentum average with the sign of $F_{s/t}^{e/o}$ being incorporated.

\subsubsection{Numerical results}\label{sec:zeemanresults}

\begin{figure}[t]
\includegraphics[width=1.0\linewidth]{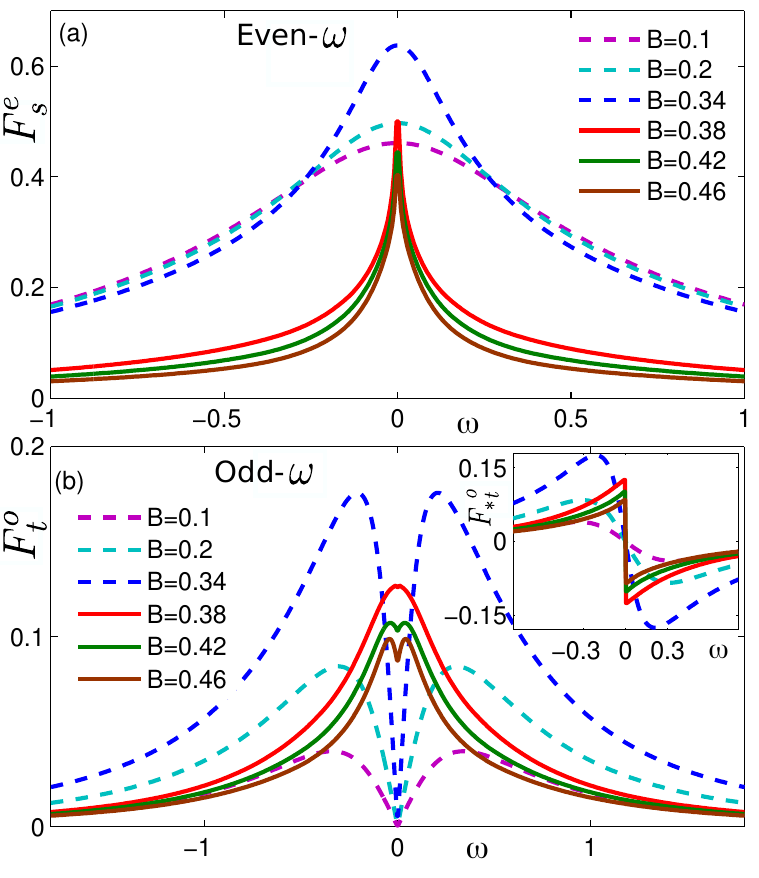} \caption{Frequency dependence of the pair correlations at different magnetic fields $B$. Momentum-averaged absolute values of the even-frequency correlations $F^e_s$ (a) and odd-frequency correlations $F^o_{t}$ (b). Inset: Momentum-averaged ${F_{*}}^o_{t}$. $B<B_{c1}=0.35$ leads to BCS phase (dashed lines) and $B>B_{c1}=0.35$ leads to FF phase (solid lines).}
\label{fig:freqfig} 
\end{figure}

Based on the theoretical framework developed in Sec.~\ref{sec:pctf}, we now evaluate the SC pair correlations numerically. We first show the frequency dependence of the even- and odd-frequency correlations for different Zeeman fields $B$ in Fig.~\ref{fig:freqfig}. Fig.~\ref{fig:freqfig}(a) shows the even-frequency correlations $F^{e}_{s}(i\omega)$ for both $B<B_{c1}$ (dashed lines) in the BCS phase and $B>B_{c1}$ (solid lines) in the FF phase. In the BCS phase, $F^{e}_{s}(i\omega)$ has a broad Gaussian-like frequency distribution with its maximum $F^{e}_{\rm{max}}$ at $\omega=0$. The BCS to FF transition at $B_{c1}$ results in a sharp change in the width of the distribution, since $F^{e}_{s}(i\omega)$ has a sharp peak at $\omega=0$ and decays much faster with $\omega$ in the FF phase than in the BCS phase. As seen in Fig.~\ref{fig:freqfig}(b), the odd-frequency correlations $F^{o}_{t}(i\omega)$ also show dramatic change in the frequency dependence when the system goes from the BCS to the FF phase. $F^{o}_{t}(i\omega)$ has its maximum value $F^{o}_{\rm{max}}$ at a finite frequency $\omega_{\rm{max}}$. Within each of the BCS phase and the FF phase, $\omega_{\rm{max}}$ changes minimally. However, $\omega_{\rm{max}}$ shows a sudden jump towards zero at $B_{c1}$. 
 The difference in the frequency dependence of odd-frequency correlations in the BCS and the FF phases becomes even more apparent in the inset of Fig.~\ref{fig:freqfig}(b), where we plot ${F_{*}}_{t}^{o}(i\omega)$ which is the total momentum sum considering also the sign of $F^{o}_{t}(k,i\omega)$, as defined in Eq.~\eqref{eq:EO_ksum}. In the BCS phase, ${F_{*}}^{o}_{t}(i\omega)$ has a smooth transition from negative to positive $\omega$. In contrast, in the FF phase ${F_{*}}^{o}_{t}(i\omega)$ has a discontinuity at $\omega=0$, which indicates a $1/\omega$ frequency dependence. We here note that one of the defining feature of $F^{o}_{t}(i\omega)$, i.e.~that it is zero at $\omega=0$, is not visible in $F^{o}_{t}(i\omega)$ in the main panel, as we do not consider the $\omega=0$ precisely.
 An important distinction has to be made at this point: $\omega_{\rm{max}}$ marking the maximum of the SC pair correlations should not be confused with the energy difference of electrons forming the finite-energy pairs. Still, the energy difference of the finite-energy pairs can be estimated by looking at the maximum values $F^{o}_{\rm{max}}$, as we show next.
 
\begin{figure}[t]
\includegraphics[width=1.0\linewidth]{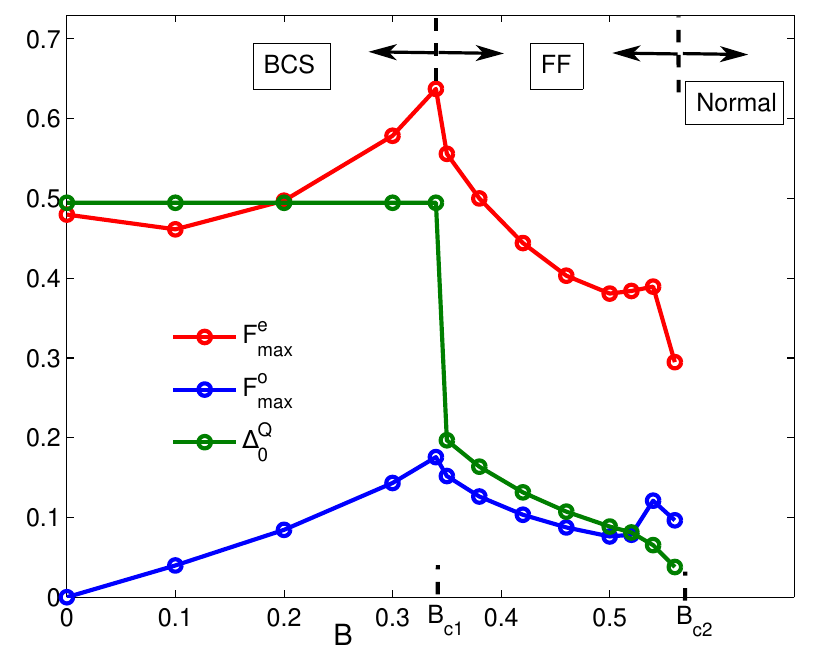} \caption{Maximum values of even- and odd-frequency correlations, $F^{e}_{\rm{max}}$ and $F^{o}_{\rm{max}}$, obtained from the main panels of Fig.~\ref{fig:freqfig} as a function of magnetic field $B$. Note that $F^{e}_{\rm{max}}$ and $F^{o}_{\rm{max}}$ occur at different frequencies. Ground state values of the SC order parameter $\Delta^{Q}_0$ is also plotted for comparison.}
\label{fig:pd} 
\end{figure}

Next we focus on the maximum values of the even- and odd-frequency correlations and relate them to the formation of finite-energy pairs. In Fig.~\ref{fig:pd} we plot the magnetic field dependence of $F^{e}_{\rm{max}}$ and $F^{o}_{\rm{max}}$, as well as the SC order parameter $\Delta^{Q}_0$ since both the even- and odd-frequency correlations depend on $\Delta^{Q}_0$, as is evident from Eqs.~\eqref{eq:feven}-\eqref{eq:fodd}. First we note that $F^{o}_{\rm{max}}$ is zero at $B=0$, verifying that odd-frequency correlations are absent in the absence of magnetic field. In contrast, $F^{e}_{\rm{max}}$ is finite for $B=0$, again verifying that even-frequency correlations exist also in the absence of finite-energy pairs. In fact, $F^{e}_{\rm{max}}$ nearly equals $\Delta^{Q}_0$ for $B=0$, showing that there is no distinction between the SC order parameter $\Delta^{Q}_0$ and the (even-frequency) SC pair correlations at $B=0$. 
With increasing $B$, both $F^{o}_{\rm{max}}$ and $F^{e}_{\rm{max}}$ increase throughout the BCS phase. This increase conforms with the increase in the energy difference of the Cooper pairs $\xi_{k\uparrow}-\xi_{-k\downarrow}=2B$ with increasing $B$, given $\Delta^{Q}_0$ is constant in the BCS phase. Although both $F^{e}_{\rm{max}}$ and $F^{o}_{\rm{max}}$ increase with increasing $B$, the absence of only $F^{o}_{\rm{max}}$ at $B=0$ clearly shows that $F^{o}_{\rm{max}}$ and hence odd-frequency correlations necessarily need finite-energy pairs. This result numerically validate our earlier analytical findings in Sec.~\ref{sec:pctf} that odd-frequency correlations are directly related to finite-energy pairs and answers our third question posed in Sec.~\ref{sec:Intro}.

With further increasing $B$, the FF phase forms where Cooper pairs obtain a finite-momentum $Q$. Here we can start addressing the first question asked in Sec.~\ref{sec:Intro}; whether finite-energy pairs also exist in the finite-momentum FF phase. In the FF phase, $(k+Q/2,\uparrow)$ electrons pair with $(-k+Q/2,\downarrow)$ electrons. The energy difference between the pairing electrons is then given by $\xi_{k+Q/2\uparrow}- \xi_{-k+Q/2\downarrow}$. In 2D, there are only certain points of the Brillouin zone where $\xi_{k+Q/2\uparrow}- \xi_{-k+Q/2\downarrow}=0$, but mostly often it is not. As shown in Eq.~\ref{eq:fodd}, $F^{o}_{k,-k}(i\omega)\propto \xi_{k+Q/2\uparrow}- \xi_{-k+Q/2\downarrow}$. Hence, showing $F^{o}_{\rm{max}}\ne 0$ is actually enough to prove that the finite-energy pairs exist in the FF phase. As seen in Fig.~\ref{fig:pd}, clearly $F^{o}_{\rm{max}}\ne 0$ in the FF phase, which proves the presence of finite-energy pairs. We here note that both $F^{o}_{\rm{max}}$ and $F^{e}_{\rm{max}}$ decrease with increasing $B$ in the FF phase. This reduction is due to the decrease of $\Delta^{Q}_0$ with increasing $B$, since $F^{e}_{\rm{max}}$ and $F^{o}_{\rm{max}}$ are both directly proportional to $\Delta^{Q}_0$, see Eqs.~\eqref{eq:feven}-\eqref{eq:fodd}. Finally, we also note that close to the normal state at $B=B_{c2}$, there is a slight increase in both $F^{o}_{\rm{max}}$ and $F^{e}_{\rm{max}}$. We attribute this to a numerical anomaly near $B=B_{c2}$ due to a large $Q$ (which increases with $B$) and small $\Delta^{Q}_0$. For fields beyond $B=B_{c2}$, the system is in the normal phase with no SC correlations.

\begin{figure}[t]
\includegraphics[width=1.0\linewidth]{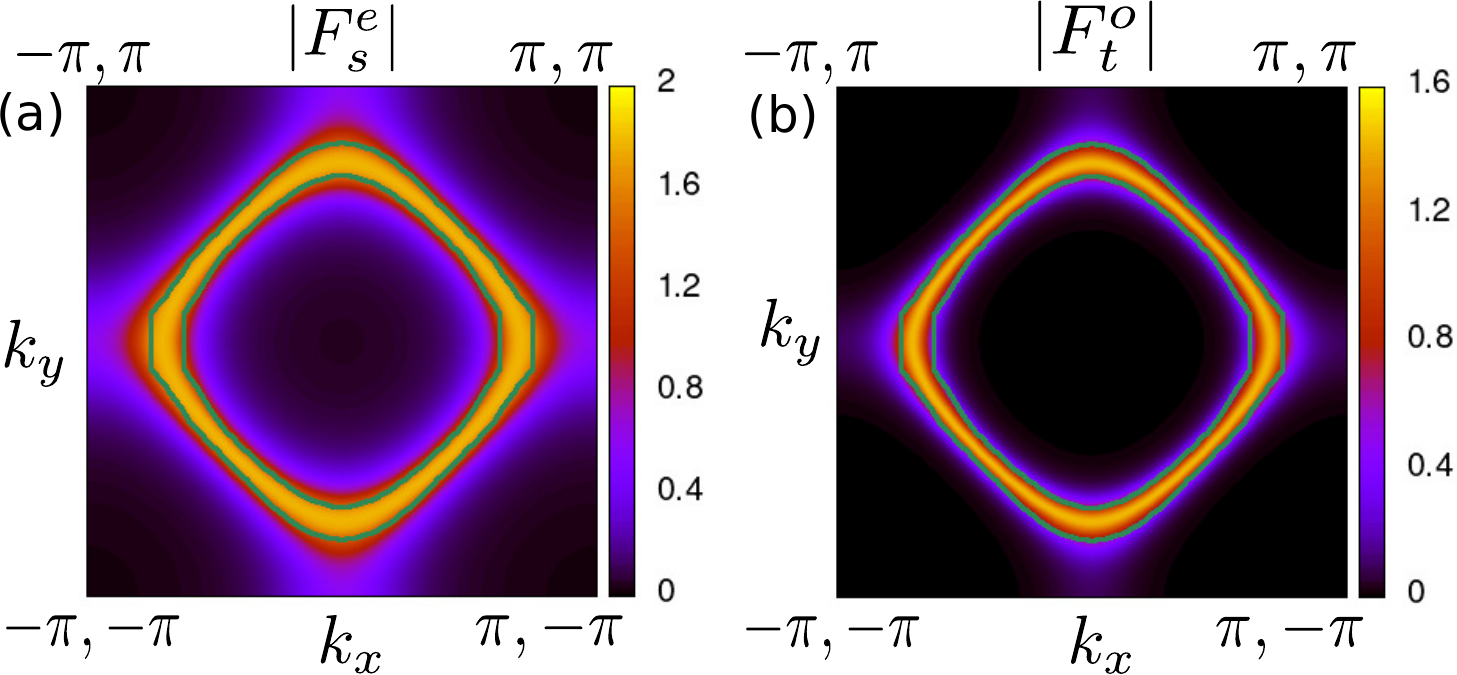} \caption{Color density map of the absolute value of the even-frequency $\left|F_{s}^{e}\right|$ (a) and odd-frequency $\left|F_{t}^{o}\right|$ (b) pair correlations in the first Brillouin zone at a fixed $\omega=0.2$ in the BCS phase ($B=0.34$). Green lines are the Fermi surfaces of $\uparrow$ and $\downarrow$ spins, i.e.~ the contours $\xi_{k\uparrow}=0$ and $\xi_{-k\uparrow}=0$.}
\label{fig:momenzee_BCS} 
\end{figure}

A more detailed understanding of the relation between the finite-energy pairs and the SC pair correlations can be established through looking at the momentum space structure of the SC correlations. We first look at the momentum-resolved SC correlations in the BCS phase. In the BCS phase for $B>0$, all the SC correlations are due to finite-energy pairs as $\xi_{k\uparrow}\ne\xi_{-k\downarrow}$. Hence, both even- and odd-frequency correlations are necessarily coming from the finite-energy pairs. In Fig.~\ref{fig:momenzee_BCS}, we show the momentum-resolved SC correlations $\left|F_{s}^{e}\right|$ (Eq.~\ref{eq:evensing}) and $\left|F_{t}^{o}\right|$ (Eq.~\ref{eq:oddtrip}) for a fixed frequency and fixed $B=0.34 < B_{c1}$. In BCS theory, the Cooper pairs are formed of the states close to the Fermi surface. Thus, we also display the Fermi surface of $\uparrow$- and $\downarrow$-spins, i.e.~the contours of $\xi_{k\uparrow}=0$ and $\xi_{k\downarrow}=0$, as green lines in Fig.~\ref{fig:momenzee_BCS}. We find that both $\left|F_{s}^{e}\right|$ and $\left|F_{t}^{o}\right|$ are mainly restricted to the $k$ points bounded by the $\xi_{k\uparrow}=0$ and $\xi_{k\downarrow}=0$ contours. 

\begin{figure}[t]
\includegraphics[width=1.0\linewidth]{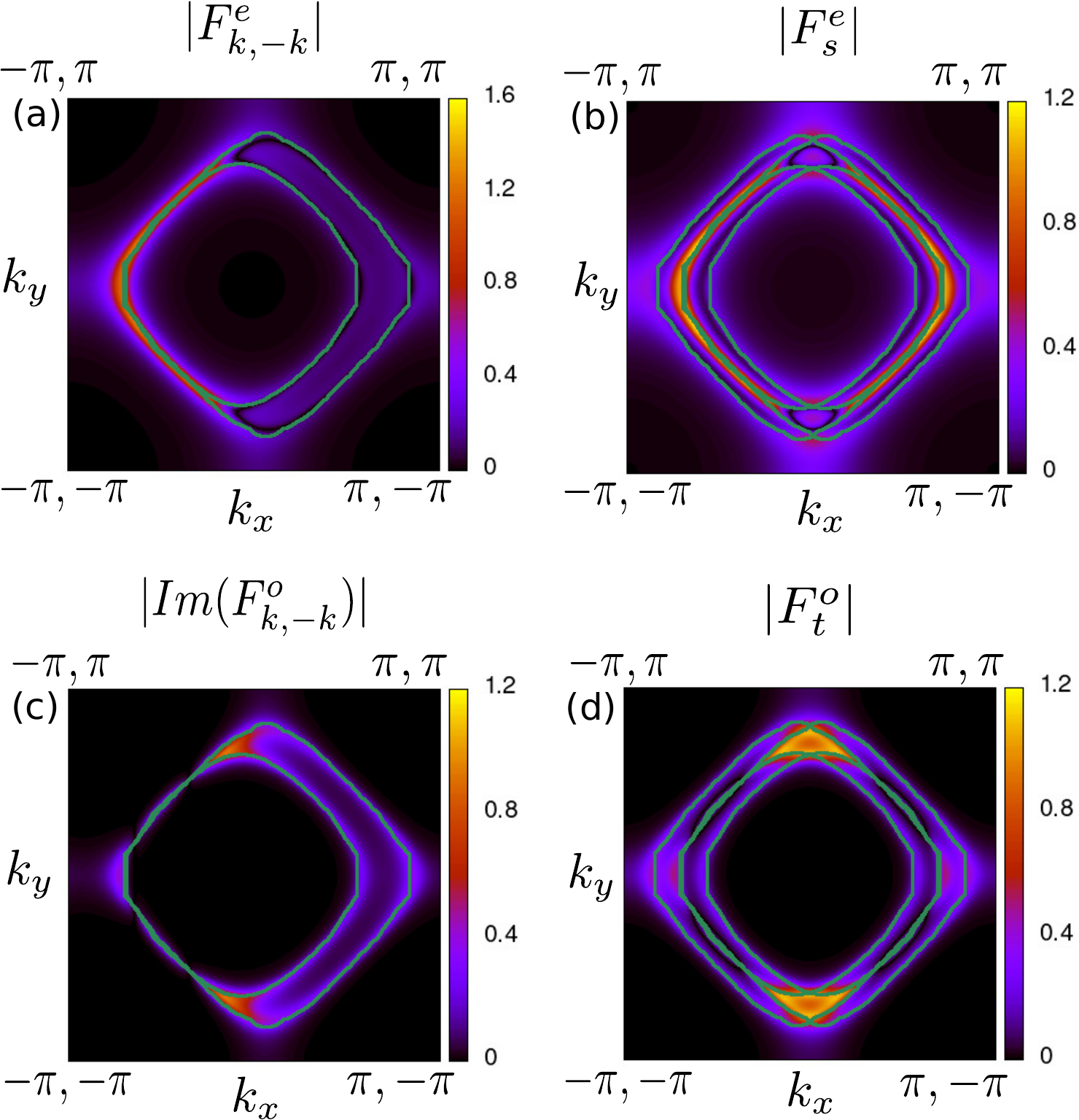} \caption{Color density map of the absolute value of the pair correlations in the first Brillouin zone at a fixed $\omega=0.2$ in the FF phase ($B=0.38$) with optimal $Q=(0.43,0)$. Absolute value of $F_{k,-k}^{e}$ (a) and total spin-singlet contribution of the even-frequency correlations (Eq.~\eqref{eq:evensing}) (b). Absolute value of $Im(F_{k,-k}^{o})$ (c) and the total spin-triplet contribution of the odd-frequency correlations (Eq.~\eqref{eq:oddtrip}) (d). Green lines in (a,c) show the contours of $\xi_{k+Q/2\uparrow}=0$ and $\xi_{-k+Q/2\downarrow}=0$. Green lines in (b,d) additionally show the contours $\xi_{k+Q/2\downarrow}=0$ and $\xi_{-k+Q/2\uparrow}=0$.}
\label{fig:momenzee_FF} 
\end{figure}

We next turn our attention to the FF phase and its momentum-resolved pair correlations. Here the two individual components $F^{e/o}_{k,-k}$ and $F^{e/o}_{-k,k}$ in Eqs.~\eqref{eq:evensing}-\eqref{eq:oddtrip} can peak at different parts of the Brillouin zone due to the presence of a finite $Q$, in contrast to the BCS phase where they always peak at the same regions. However, $F^{e/o}_{k,-k}$ and $F^{e/o}_{-k,k}$ are still mirror reflections of each other about the $k_x=0$ line, as $Q$ is only along the $x$-direction. Hence, to get a comprehensive picture, we plot in Fig.~\ref{fig:momenzee_FF} both the total contributions $F_{s/t}^{e/o}$ in (b,d) together with one  of the individual component $F^{e/o}_{k,-k}$ in (a,c) for a fixed frequency $\omega=0.2$ where odd-frequency correlations are considerable. There is also another important distinction between the BCS phase and the FF phase. In the BCS phase, the SC correlations are restricted between the contours of the Fermi surfaces $\xi_{k\uparrow/\downarrow}=0$, as also seen in Fig.~\ref{fig:momenzee_BCS}. However, in the FF phase, $(k+Q/2,\uparrow)$ electrons pair with $(-k+Q/2,\downarrow)$ electrons and as a result we expect the SC correlations to be dominant near $\xi_{k+Q/2\uparrow}=0$ and $\xi_{-k+Q/2\downarrow}=0$, instead of $\xi_{k\uparrow/\downarrow}=0$. Hence in Fig.~\ref{fig:momenzee_FF} it is most illustrative to overlay the contours of $\xi_{k+Q/2\uparrow}=0$ and $\xi_{-k+Q/2\downarrow}=0$ as green lines. As seen in (a) and (c), these two green lines are shifted in the $x$-direction due to the uniaxial nature of the optimum $Q$ and only nearly merge  for a line of $k$-points in a  region $k_x<0$ satisfying $\xi_{k+Q/2\uparrow} \approx \xi_{-k+Q/2\downarrow}$. As a result, at these points there can be no finite-energy pairs. As seen in Fig.~\ref{fig:momenzee_FF}(a) $F_{k,-k}^{e}$ has a clear maximum exactly in this region where $\xi_{k+Q/2\uparrow} \approx \xi_{-k+Q/2\downarrow}$. This establishes that the total even-frequency correlations, as also seen in (b), are largely dominated by zero-energy pairs. In contrast, we show in Fig.~\ref{fig:momenzee_FF}(c) that $F_{k,-k}^{o}$ is zero in this region where $\xi_{k+Q/2\uparrow} \approx \xi_{-k+Q/2\downarrow}$. Instead, we find that $F_{k,-k}^{o}$ is maximum in regions where the two green lines start deviating from each other, i.e.~in the regions where necessarily $\xi_{k+Q/2\uparrow} \ne \xi_{-k+Q/2\downarrow}$ but still with a proximity to low-energy excitations indicated by $\xi_{k+Q/2\uparrow}=0$ and $\xi_{-k+Q/2\downarrow}=0$. Thus, these momentum space findings show clearly that odd-frequency correlations are only present for finite-energy pairs, where $\xi_{k+Q/2\uparrow} \ne \xi_{-k+Q/2\downarrow}$, while even-frequency correlations are mainly driven by zero-energy pairs where $\xi_{k+Q/2\uparrow} \approx \xi_{-k+Q/2\downarrow}$. The results of Fig.~\ref{fig:momenzee_FF} are qualitatively similar for frequencies other than $\omega=0.2$, except for very low $\omega$ where even $F_{k,-k}^{e}$ is also generated by finite-energy pairs.

To summarize, in this section we provide answers to the first and the third questions posed in Sec.~\ref{sec:Intro}. In particular, we find that the odd-frequency SC correlations are directly related to finite-energy pairs by showing both momentum-averaged and momentum-resolved correlations, while such a relation is found to be absent for the even-frequency SC correlations for most frequencies. Further, by using the unique analytical relation of odd-frequency correlations and finite-energy pairs in Eq.~\eqref{eq:fodd}, we show that finite-energy pairs are also present in a finite-momentum phase formed under applied magnetic field. Thus, the system relaxing into a finite-momentum FF state with increasing magnetic field, does not remove the finite-energy pairs nor the odd-frequency pairing.

\section{Finite-energy pairing in the absence of magnetic field}\label{sec:pairhopping}
Having established the existence of finite-energy and odd-frequency SC pairs in a finite-momentum FF state driven by an magnetic field in the previous Section, we next aim to generalize these results by studying a finite-momentum state without an applied magnetic field. 
As already explained in Sec.~\ref{sec:Intro}, in the absence of magnetic field or, equivalently, in the absence of any spin-splitting of the Fermi surface, any finite-momentum spin-singlet pairing state necessarily consists of finite-energy pairs. The aim here is therefore to answer the second question posed in Sec.~\ref{sec:Intro}, i.e.~establish the nature of finite-energy pairs in a finite-momentum phase generated spontaneously in the absence of any applied magnetic field.
 In order to investigate such a finite-momentum state, we first need to set up a viable model where a finite-momentum phase can be spontaneous generated.

\subsection{Model and ground state}\label{sec:modelpair}

Finding microscopic models giving rise to a finite-momentum SC state in the absence of magnetic fields has been challenging. In the literature, most works have been focused on the $d$-wave cuprate superconductors, where finite-momentum superconductivity seems to explain several mysterious experimental findings \cite{Berg09}. In the context of the cuprate superconductors, finite-momentum superconductivity is often referred to as a pair density wave state where the superconducting order parameter modulates in real space even in the absence of any applied magnetic field \cite{Agterberg20}. A pair density wave state can generate secondary charge density wave modulations with a wave vector twice of the superconducting modulation wave vector. However, a charge density wave state, also discussed in the context of cuprate superconductors \cite{Comin16}, is distinct from a pair density wave state since it does not necessarily lead to modulating superconducting order. Here, we discuss the microscopic model pertaining to a pair density wave like finite-momentum SC state and not a charge density wave state. It has been shown that a finite-momentum SC state can be obtained in a real space $t$-$J$ Hamiltonian appropriate as a low-energy description for the high-temperature cuprate superconductors, but only at very strong interactions strengths \cite{Loder10,Waardh17}. In fact, the minimum interaction strength required to obtain a ground state with finite-momentum superconductivity in this model has been shown to be six times the hopping amplitude, which is usually considered to be too large \cite{Waardh17}. However, more recently, it has been shown that a finite-momentum SC state can in fact be obtained with more reasonable interaction strengths, of the order of the hopping amplitude, by considering a Hamiltonian with a real space nearest-neighbor attraction as in the $t$-$J$ model augmented with periodically modulated longer range pair hopping \cite{Waardh17}. Such pair hopping terms have quite often been proposed in Hubbard-like models in different systems, including multi-orbital systems\cite{Herbrych18,Japaridze01}. In the context of cuprates, this pair hopping can be directly motivated from the Josephson coupling in the $\pi$-junctions formed near stripe domain walls \cite{Berg09}. The range and the period of the modulation in the pair hopping interaction are then the same as the experimentally observed stripe periods in the cuprates \cite{Tranquada20}. The resulting pair hopping (PH) model on a 2D square lattice in momentum space can be written as,
\begin{eqnarray}
&&H_{\rm {PH}}=\sum_{k,\sigma} \xi_{k} c_{k \sigma}^{\dagger} c_{k \sigma} \nonumber \\
&&+ \sum_{k,k',q} V_{k,k^{\prime},q} c_{k+q/2 \uparrow}^{\dagger} c_{-k+q/2 \downarrow}^{\dagger}c_{-k'+q/2 \downarrow}c_{k'+q/2 \uparrow}, \nonumber  \\
&& \label{eq:phHamil}
\end{eqnarray}
where the electron dispersion is now $\xi_{k}=-2t(\cos(k_x)+\cos(k_y))-4t'\cos(k_x)\cos(k_y)-\mu$ with $t=1$ still  the energy unit and an additional next nearest neighbor hopping $t'=-0.3$ to mimic a prototype cuprate band structure\cite{Norman07}. We further tune $\mu$ such that the average density of electrons $\rho=\sum_{k,\sigma}\langle c^{\dagger}_{k\sigma}c_{k\sigma} \rangle$  is fixed to $0.65$, a value that is already known to favor the finite-momentum state \cite{Waardh17}. Following Ref.~\onlinecite{Waardh17}, the pair hopping interaction coming from a nearest-neighbor attraction is given by,
\begin{equation}
V_{k,k^{\prime},q}=-V\Gamma(q)\left(\gamma(k)\gamma(k')+\eta(k)\eta(k')\right), \label{eq:phint} 
\end{equation}
where $\gamma(k)=\cos(k_x)+\cos(k_y)$ and $\eta(k)=\cos(k_x)-\cos(k_y)$ are the two form factors for nearest-neighbor attraction on a square lattice, and $V$ is the constant attraction strength. The aspect of longer range pair hopping is embedded in the factor $\Gamma(q)$ given by,
\begin{equation}
\Gamma(q)= e^{-\frac{\left(q_x-\tilde{Q}\right)^2}{2\kappa_x^2}}+e^{-\frac{\left(q_x+\tilde{Q}\right)^2}{2\kappa_x^2}}, 
\label{eq:gamma}
\end{equation}
where $\kappa_{x}$ denotes the range of the hopping with the limit $\kappa_{x}\rightarrow \infty$ meaning nearest-neighbor pair hopping. The modulation $\tilde{Q}=2\pi/P\hat{x}$ is here introduced to mimic the presence of stripe modulation, with $P=8$ taken from experimental motivation of the modulation wave vector of stripes in cuprates \cite{Tranquada20}. The choice of $\tilde{Q}$ breaks the rotational symmetry since the modulation is only along $x$-direction. In order to have only finite-momentum pairing the hopping range has to be smaller than the modulation wave vector \cite{Waardh17} and thus we here consider $\kappa_x=0.2$. A mean-field decomposition of the Hamiltonian in Eq.~\eqref{eq:phHamil} in the Cooper channel results in a similar Hamiltonian as in Eq.~\eqref{eq:Hamil}, but now with zero magnetic field and given by, 
\begin{eqnarray}
H_{\rm {PH}}&=&\sum_{k,\sigma} \xi_{k} c_{k \sigma}^{\dagger} c_{k \sigma} + \sum_{k} \left( \Delta^{Q}_{k} c_{-k+Q/2 \downarrow} c_{k+Q/2 \uparrow} + \textrm{H.c.} \right) \nonumber \\
&&+ \text{constant},
\label{eq:phHamilmf}
\end{eqnarray}
where $\Delta^{Q}_{k}$ is the spin-singlet SC order parameter obtained by the self-consistency relation,
\begin{equation}
\Delta^{Q}_k=\sum_{k^{\prime}}V_{k,k^{\prime},Q} \langle c_{k^{\prime}+Q/2 \uparrow}^{\dagger} c_{-k^{\prime}+Q/2 \downarrow}^{\dagger} \rangle.
\label{eq:scph}
\end{equation}
This self-consistency relation is different from Eq.~\eqref{eq:sc} because of the explicit $Q$ dependence in $V_{k,k^{\prime},Q}$, which favors finite-momentum superconductivity, whereas the self-consistency relation in Eq.~\eqref{eq:sc} does not need an explicit $Q$ dependence in $V_{k,k^{\prime}}$ due to the presence of the external magnetic field. Furthermore, $V_{k,k^{\prime},Q}$ also has an explicit $k,k^{\prime}$ momentum dependence as seen in Eq.~\eqref{eq:phint}. Due to the $\eta(k)$ and $\gamma(k)$ components in $V_{k,k^{\prime},Q}$, we can decompose $\Delta^{Q}_k$ as $\Delta^{Q}_k=\Delta^{Q}_d\eta(k)+\Delta^{Q}_s\gamma(k)$, with $\Delta^{Q}_d$ being the $d$-wave SC order parameter and $\Delta^{Q}_s$ being the extended $s$-wave SC order parameter\cite{SudboBook}. Below we show results for $V=1.3$, a value we have checked to be close to the minimum interaction strength required to obtain finite $\Delta^{Q}_k$. We have also used other $V$, finding no qualitative difference. We further work with the same system size as in Sec.~\ref{sec:modelZeeman}. We note here that the effects of strong electronic correlations in cuprate superconductors are not explicitly considered in this work. This is a reasonable approximation since we use a relatively small $\rho=0.65$. When approaching half-filling, the role of strong correlations is expected to increase. A finite-momentum superconducting state obtained in the presence of magnetic field has been shown to be stable when strong correlations are incorporated within the Gutzwiller approximation \cite{Maska10}. We expect similar stability of the finite-momentum state even in the absence of magnetic field and hence the strong correlation effects present if moving closer to half-filling will likely not change our results.

\begin{figure}[t]
\includegraphics[width=0.7\linewidth]{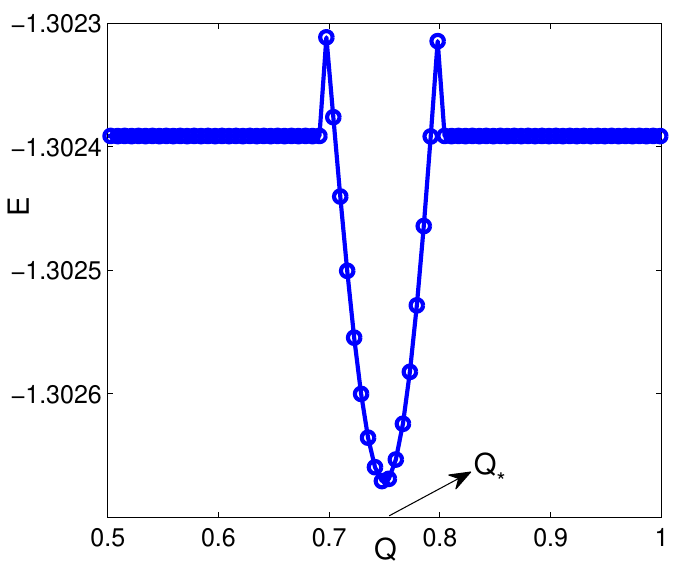} \caption{Ground state energy as a function of $Q$ in the pair hopping model Eq.~\eqref{eq:phHamilmf}. $Q$ is taken to be along the $x$-axis and $Q_y=0$. Minimum in  energy is identified with arrow and corresponds to the optimal $Q_{*}$.}
\label{fig:energyQ} 
\end{figure}

As in Sec.~\ref{sec:modelZeeman}, the Hamiltonian in Eq.~\eqref{eq:phHamilmf} can be written in a matrix form $\hat{H}_{\rm {PH}}$ similar to Eq.~\eqref{eq:Hamilmat}, but now the diagonal terms do not have a spin index, i.e.~$\xi_{k+Q/2\uparrow}$ will be replaced by $\xi_{k+Q/2}$ and $\xi_{-k+Q/2\downarrow}$ by $\xi_{-k+Q/2}$. We can then follow the same self-consistency procedure discussed in Sec.~\ref{sec:modelZeeman}, but now with two self-consistent order parameters $\Delta^{Q}_{d}$ and $\Delta^{Q}_{s}$, instead of only one $\Delta^{Q}_0$. In order to obtain the global energy minimum, we here calculate the ground state energy using $E=\sum_{k,\sigma}\xi_{k}\langle c^{\dagger}_{k\sigma}c_{k\sigma} \rangle-(\Delta^{Q}_d)^2/(V\Gamma(Q))-(\Delta^{Q}_s)^2/(V\Gamma(Q))+\mu\rho$ \cite{Waardh17}. This expression for ground state energy is equivalent to the one used in Sec.~\ref{sec:modelZeeman} after appropriately including the momentum- and modulation-dependent interaction strength given in Eq.~\eqref{eq:phint}. We find that the self-consistent value of $\Delta^{Q}_{s}$ is very small for all $Q$ and negligible compared to $\Delta^{Q}_{d}$. In Fig.~\ref{fig:energyQ} we show the variation of $E$ with $Q$. We have considered only uniaxial values of $Q$ along the $x$-axis to find the global energy minimum, since $\tilde{Q}$ in Eq.~\eqref{eq:gamma} is only along $x$-direction. As seen in Fig.~\ref{fig:energyQ}, $E$ forms a minima at an optimal $Q=Q_*$. This obtained $Q_*$ matches well with the findings of Ref.~\onlinecite{Waardh17}. This result establishes that it is possible to find a finite-momentum SC ground state with the pair hopping Hamiltonian $H_{\rm PH}$ for the parameters considered in this work. Note that,  again, since we only consider one $Q$ and consequently, the finite-momentum SC ground state correspond to the FF state but now in the absence of a magnetic field.

\subsection{Superconducting correlations}\label{sec:correlationspair}

After finding the optimal value of $Q=Q_*$ and the self-consistent order parameters $\Delta^{Q}_{s}$ and $\Delta^{Q}_{d}$ of the ground state, we next look at the SC correlations. Using the same procedure as in Sec.~\ref{sec:correlationsZeeman}, we first obtain the Green's function $\tilde{G}$ by inverting the $2\times 2$ matrix $\tilde{G}^{-1}(i\omega)=i\omega-\hat{H}_{\rm {PH}}$ (in this Section we use tilde to indicate quantities for the PH model). The pair correlator $\tilde{F}_{k,-k}(i\omega)$ is then given by the off-diagonal elements of the Green's function, $\tilde{G}_{12}(i\omega)$. Since $\hat{H}_{\rm {PH}}$ and $\hat{H}_{B}$ have same matrix structure, $\tilde{F}_{k,-k}(i\omega)$ can also be decomposed into even-frequency $\tilde{F}^{e}_{k,-k}(i\omega)$ and odd-frequency $\tilde{F}^{o}_{k,-k}(i\omega)$ components and we find analytically,
\begin{eqnarray}
&&\tilde{F}^{e}_{k,-k}(i\omega)=\frac{-\Delta^{Q}_k\left( \xi_{k+Q/2}\xi_{-k+Q/2}+(\Delta^{Q}_k)^2+\omega^2 \right)}{\tilde{D}},\label{eq:fevenph}\\
&&\tilde{F}^{o}_{k,-k}(i\omega)=\frac{i\omega\Delta^{Q}_k\left( \xi_{k+Q/2}-\xi_{-k+Q/2} \right)}{\tilde{D}},\label{eq:foddph}
\end{eqnarray}
where
\begin{eqnarray}
&&\tilde{D}=\left( \xi_{k+Q/2}\xi_{-k+Q/2}+(\Delta^{Q}_k)^2+\omega^2 \right)^2 \nonumber \\
&&+\omega^2\left( \xi_{k+Q/2}-\xi_{-k+Q/2} \right)^2. \label{eq:Dph}
\end{eqnarray}
The functional forms of $\tilde{F}^{e}_{k,-k}$, $\tilde{F}^{o}_{k,-k}$, and $\tilde{D}$ are the same as in Eqs.~\eqref{eq:feven}-\eqref{eq:D}, only with the spin labels of $\xi_{k}$ being removed and $\Delta^{Q}_k$ with a momentum dependence. Thus, again, the odd-frequency correlations, $\tilde{F}^{o}_{k,-k}$, is directly proportional to the energy difference, $\xi_{k+Q/2}-\xi_{-k+Q/2}$, of the electrons forming the Cooper pairs.

The spin symmetries of the SC pair correlations also follow the same analysis as in Sec.~\ref{sec:correlationsZeeman}. Noticing that under momentum exchange $\tilde{F}^{e}_{-k,k}=\tilde{F}^{e}_{k,-k}$ and $\tilde{F}^{o}_{-k,k}=-\tilde{F}^{o}_{k,-k}$, and from Eqs~\eqref{eq:evensing}-\eqref{eq:oddtrip}, we find that only the spin-singlet components $\tilde{F}^{e/o}_{s}(k,i\omega)$ of both odd- and even-frequency correlations persist. This absence of spin-triplet components is a consequence of the net spin polarization being zero in the absence of an applied magnetic field. This is an important distinction between the FF phase obtained here and the magnetic-field induced FF phase studied in Sec.~\ref{sec:modelZeeman} where both spin-singlet and -triplet components are present. We here also define the momentum-averaged absolute values of even- and odd-frequency correlations $\tilde{F}^{e/o}_{s}(i\omega)$ in the same way as in Eq.~\eqref{eq:EO_modksum}. We here do not show the momentum sums keeping the signs of $\tilde{F}^{e/o}_{s}(k,i\omega)$ as in Eq.~\eqref{eq:EO_ksum}, since we find it to give very similar behavior to the momentum-averaged results.

\begin{figure}[t]
\includegraphics[width=1.0\linewidth]{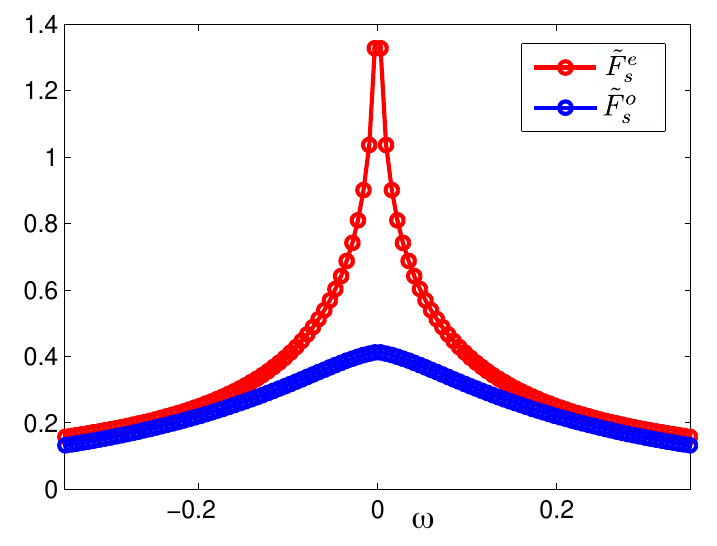} \caption{Frequency dependence of the pair correlations in the pair hopping model, similar to Fig.~\ref{fig:freqfig} but here in the absence of magnetic field. Momentum-averaged absolute values of the even- and odd-frequency correlations, $\tilde{F}^{e}_{s}$ and $\tilde{F}^{o}_{s}$, respectively.}
\label{fig:evenoddph} 
\end{figure}

To be able to gain detailed understanding we numerically evaluate the SC correlations in Eqs.~\eqref{eq:fevenph}-\eqref{eq:Dph}. In Fig.~\ref{fig:evenoddph} we show the frequency dependence of the momentum-averaged absolute values, $\tilde{F}^{e}_{s}$ and $\tilde{F}^{o}_{s}$. We find this frequency dependence to be quite similar to the ones obtained in the FF phase in the presence of magnetic field (solid lines in Fig.~\ref{fig:freqfig}). In particular, $\tilde{F}^{o}_{s}$ is generally finite and has its maximum at a frequency $\omega$ very close to zero. 
Since the odd-frequency correlations are directly proportional to the finite-energy pairs as seen in Eq.~\eqref{eq:foddph}, this result directly verify the presence of finite-energy pairs.

\begin{figure}[t]
\includegraphics[width=1.0\linewidth]{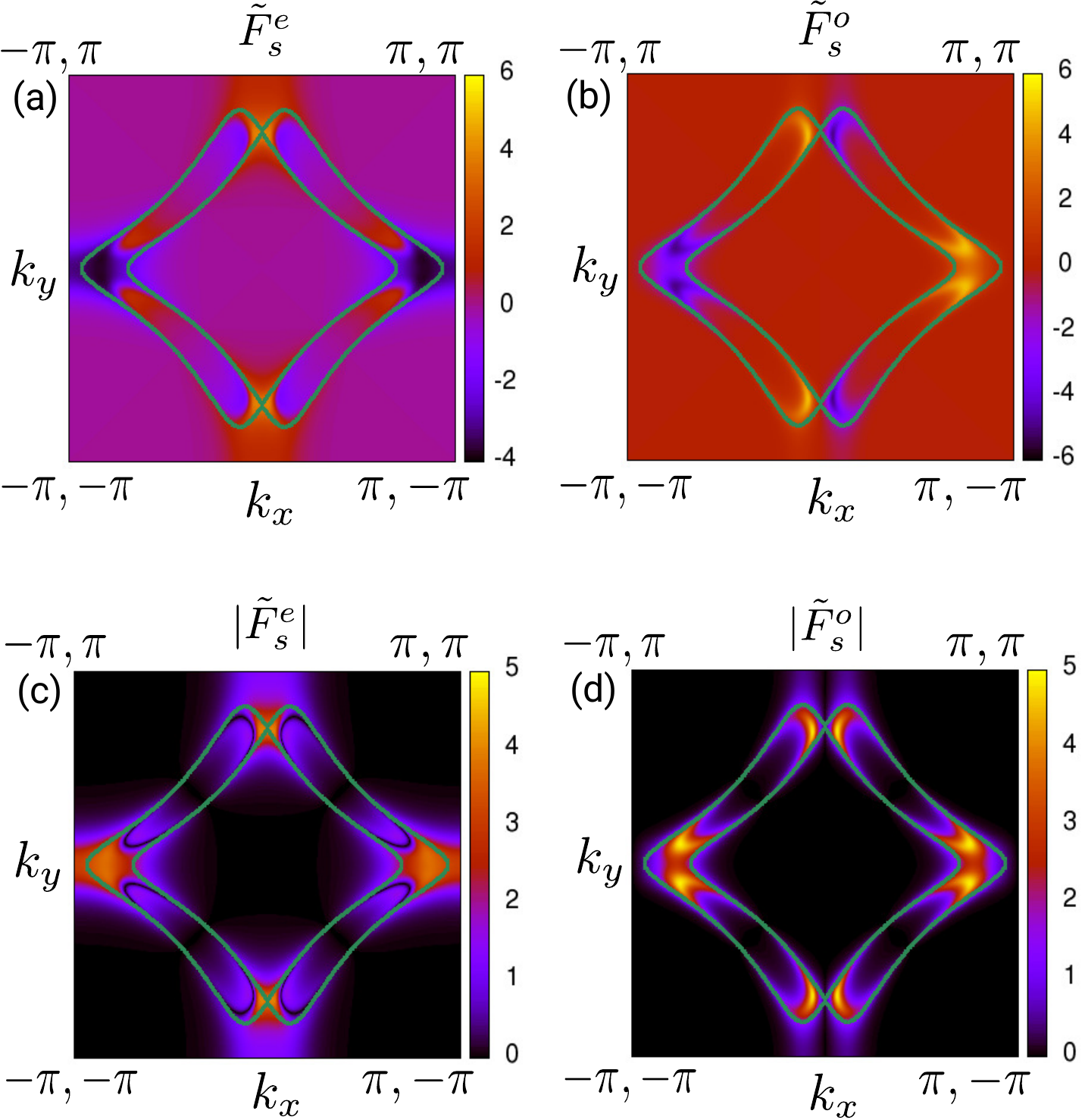} \caption{Color density map of the even-frequency (a,c) and odd-frequency (b,d) correlations in the first Brillouin zone at a fixed $\omega=0.1$ in the pair hopping model for optimal $Q=(0.75,0)$. Lower plots show absolute values of upper row. Green lines show the contours of $\xi_{k+Q/2}=0$ and $\xi_{-k+Q/2}=0$.}
\label{fig:momenpair} 
\end{figure}

Since the ground state SC order parameter $\Delta_k$ is found to be mainly $d$-wave in nature, it generates the possibility of finding unconventional momentum space structure of the SC correlations. Hence, in Fig.~\ref{fig:momenpair}(a,b) we show as a color density plot the momentum-resolved $\tilde{F}^{e}_{s}$ and $\tilde{F}^{o}_{s}$ for a fixed frequency $\omega=0.1$ where odd-frequency correlations are considerable. As seen, $\tilde{F}^{e}_{s}$ peaks near the anti-nodal region, i.e.~in the regions around ($\pm\pi,0$) and ($0,\pm\pi$), and displays a clear $d$-wave signature with sign-changing values. In contrast, $\tilde{F}^{o}_{s}$ shows a sign-change between $+k_x$ and $-k_x$, indicating that these correlations have $p$-wave character. The $p$-wave character is also consistent with Eq.~\eqref{eq:oddsing} when using the odd-frequency SC correlations defined in Eq.~\eqref{eq:foddph}. This finding is remarkable in the sense that a $d$-wave finite-momentum SC order parameter generates significant values of $p$-wave odd-frequency pair correlations in the bulk. In the literature, $p$-wave odd-frequency correlations have predominantly only been discussed in the context of heterostructures \cite{Golubov09,Tanaka12,Lothman21}. Here, we do need any heterostructures, but the generation of the $p$-wave correlations is due to the broken spatial parity in the FF phase \cite{Waardh17}. 

Finally, in order to connect the SC correlations with finite-energy pairs, we also show the contours of $\xi_{k+Q/2}=0$ and $\xi_{-k+Q/2}=0$ as overlaid green lines in the color density map of Fig.~\ref{fig:momenpair} in the same spirit as Fig.~\ref{fig:momenzee_FF}. As seen from the plot, $\xi_{k+Q/2}$ and $\xi_{-k+Q/2}$ intersect only at two $k$-points. At these $k$-points where $\xi_{k+Q/2}=\xi_{-k+Q/2}$, $|\tilde{F}^{e}_{s}|$ in (c) is largest but $|\tilde{F}^{o}_{s}|=0$ in (d). Hence, even-frequency correlations are mainly formed by pairs close to zero energy and with no energy difference, i.e.~zero-energy pairs, whereas odd-frequency correlations requires finite-energy pairs, the same as the findings as in Sec.~\ref{sec:Zeeman}. Additionally, $|\tilde{F}^{e}_{s}|$ is also seen to be large near $k \approx (\pm \pi,0)$. Around these regions, $|\xi_{k+Q/2}-\xi_{-k+Q/2}|\approx 0$, even though $\xi_{k+Q/2}\ne 0$ and $\xi_{-k+Q/2} \ne 0$ individually. Thus, also the large values of $|\tilde{F}^{e}_{s}|$ near $k \approx (\pm \pi,0)$ are driven by zero-energy pairs. We note that similar regions with zero-energy pairs away from the green contours of $\xi_{k+Q/2}=0$ and $\xi_{-k+Q/2}=0$ does not occur in Sec.~\ref{sec:Zeeman} due to the nature of the bands considered. For frequencies other than $\omega=0.1$, the qualitative features are similar to Fig.~\ref{fig:momenpair}, except for very low $\omega$ where $|\tilde{F}^{e}_{s}|$ turns out to also be generated by finite-energy pairs. 

To summarize, our findings of this section show that a spontaneously formed FF phase in the absence of magnetic field gives a realistic example where finite-energy and finite-momentum pairs co-exist instead of competing with each other. We further show that the finite-energy pairs are also intimately connected to the odd-frequency correlations, similar to the findings of Sec.~\ref{sec:Zeeman}. We additionally find that the odd-frequency correlations have a momentum structure with orthogonal orbital symmetries compared to the underlying $d$-wave SC order parameter, which generates the correlations. Overall, this clarifies the nature of the finite-energy pairs in an FF state without an applied magnetic field and thus answers the second question posed in Sec.~\ref{sec:Intro}. Combined with the results in the previous section Sec.~\ref{sec:Zeeman}, these findings establish that finite-energy pairs exist generally in finite-momentum FF phases and that they directly generate odd-frequency SC correlations. This firmly answers the first three questions posed for this work.

\section{Meissner effect}\label{sec:meissner}

Having understood the nature of the SC correlations in different SC phases with finite-energy and finite-momentum pairs, we calculate in this section the effect of these correlations on the experimentally relevant Meissner effect in order to address the fourth and final question. We choose the Meissner effect as it is one of the defining features of a superconductor, measuring its expulsion of an external magnetic field, a so-called diamagnetic Meissner effect. However, odd-frequency correlations have historically been shown to instead produce an unusual paramagnetic Meissner effect \cite{Yokoyama11,Alidoust14,Mironov12,Bernardo15}, which would mean the superconductor attracts the magnetic field and subsequently become unstable. It is thus highly interesting to understand the Meissner effect in systems where odd-frequency correlations are strong. We here primarily discuss the Meissner effect in the conventional superconductor in an applied magnetic field as studied in Sec.~\ref{sec:Zeeman}, but comment on the results for the pair hopping model discussed in Sec.~\ref{sec:pairhopping} towards the end.

The usual diamagnetic Meissner effect correspond to a positive superfluid weight of a superconductor \cite{Tinkhambook}, whereas a paramagnetic Meissner effect would indicate a negative superfluid weight. 
Within Kubo linear response theory, the superfluid weight $D_s$ is given by \cite{Scalapino93}
\begin{equation}
D_s=\langle -k_x \rangle - \Lambda_{xx}(q_x=0,q_y \rightarrow 0,i\nu=0), \label{eq:supd}\\
\end{equation}
where we have ignored the scaling factor $e^2\pi$ to avoid dealing with very small numbers. We choose the response in the $x$-direction since the modulation wave vector in the FF state is also chosen to be in the $x$-direction and then $\langle k_x \rangle$ is the kinetic energy per site along the $x$-direction. The transverse current-current correlation function $\Lambda_{xx}$ is given by 
\begin{equation}
\Lambda_{xx}(q,i\nu)=\frac{1}{N}\int_0^{1/T}d\tau e^{i\nu\tau}\langle j_x^p(q,\tau)j_x^p(-q,0)\rangle,
\label{eq:curcur}
\end{equation}
where $N$ is the system size, $q$ is the bosonic momentum, and $\nu=2\pi mT$ ($m$ is a positive integer) is the bosonic Matsubara frequency with $T$ being the temperature. For calculating the superfluid weight in Eq.~\eqref{eq:supd}, we take the long wavelength ($q_y \rightarrow 0$) and static ($i\nu=0$) limit of $\Lambda_{xx}$, while setting $q_x=0$ since we look at the response in the $x$-direction \cite{Scalapino93}. The current-current correlation $\Lambda_{xx}$ can be calculated using the Green's function $G$ \cite{Bruusbook,Fominov15,Hoshino14,Parhizgar21},
\begin{eqnarray}
&&\Lambda_{xx}(q,i\nu) \nonumber \\
&=&-\sum_{k,i\omega} Tr\left[ G(k,i\omega)J(k)G(k+q,i\omega+i\nu)J(k+q) \right], \nonumber \\
\label{eq:curcurgr} 
\end{eqnarray}
where
\begin{equation}
J(k)=\left(\begin{array}{cc} J_1 & 0 \\
0 & J_2 \\
\end{array}\right),
\end{equation}
with $J_1=v_{k+Q/2}$, $J_2=v_{-k+Q/2}$, and $v_{k}=\partial\xi_{k}/\partial k$ \cite{Hoshino14}. Using Eqs.~\eqref{eq:supd} and \eqref{eq:curcurgr}, we arrive at a total superfluid weight given by,
\begin{eqnarray}
D_s&=&\langle -k_x \rangle+\sum_{k,i\omega} \left( J_1^2 G_{11}G_{11}+J_2^2 G_{22}G_{22}+2J_1J_2G_{12}G_{21}\right) \nonumber\\
&=&\langle -k_x \rangle+\sum_{k,i\omega} \left( J_1^2 G_{11}G_{11}+J_2^2 G_{22}G_{22} \right. \nonumber \\
&&\left.+2J_1J_2F^{e}_{k,-k}F^{e}_{k,-k}+2J_1J_2F^{o}_{k,-k}F^{o}_{k,-k}\right) \nonumber \\
&=&K_n+K_a^{e}+K_a^{o} \label{eq:totalsupd} 
\end{eqnarray}
where the first term $K^n$ isolates the contribution of the normal (diagonal) part of the Green's function and the sum of the last two terms $K_a=K_a^{e}+K_a^{o}$ identifies the contribution of the anomalous (off-diagonal) part of the Green's function, only present in the superconducting state. Here we have used the fact that contributions being products of even- and odd-frequency correlations after frequency summation identically vanish. 
\begin{figure}[t]
\includegraphics[width=1.0\linewidth]{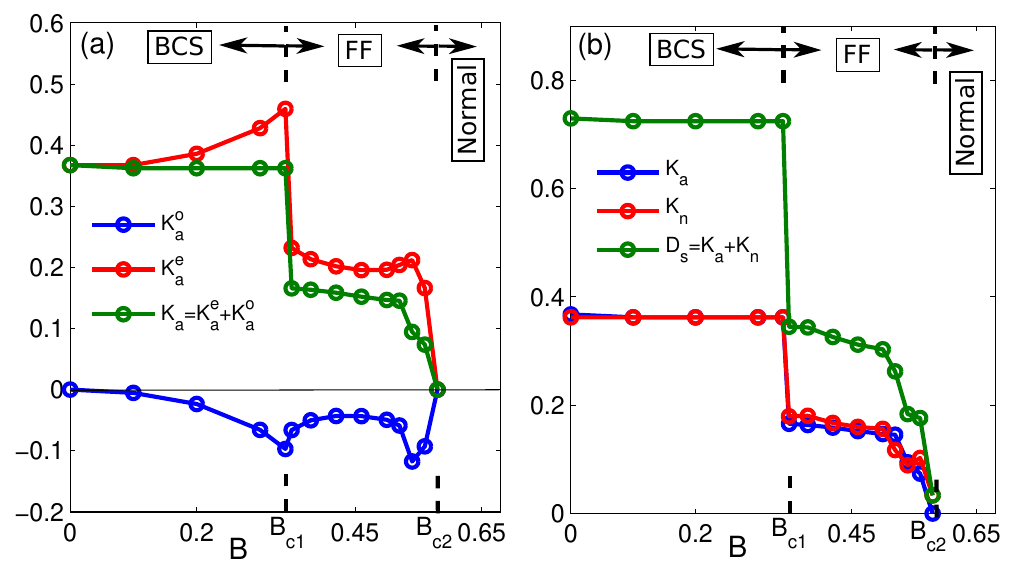} \caption{Meissner effect as a function of magnetic field $B$. (a) Contribution from the pair correlations to the superfluid weight $K_{a}$ and individual contributions from even-frequency $K_{a}^{e}$ and odd-frequency $K_{a}^{o}$ correlations. (b) Total superfluid weight $D_s$, and the contributions coming from diagonal part of the Green's function $K_n$ and pair correlations $K_a$.}
\label{fig:meis} 
\end{figure}

We use the solutions for $G_{11}$, $G_{22}$, $F^{e}_{k,-k}$, and $F^{o}_{k,-k}$  obtained in Sec.~\ref{sec:Zeeman} in Eqs.~\eqref{eq:feven}-\eqref{eq:fodd} to arrive at the superfluid weight. First, in Fig.~\ref{fig:meis}(a) we analyze the anomalous contribution $K_a$ and its even- and odd-frequency contributions, $K_a^{e}$ and $K_a^{o}$, respectively, as a function of magnetic field $B$. We find that $K_a^{e}$ gives a positive contribution to the superfluid weight, whereas $K_a^{o}$ gives a negative contribution for all $B<B_{c2}$. This means that the even-frequency pair correlations give a diamagnetic contribution to the Meissner effect, while the odd-frequency correlations give a paramagnetic contribution. $K_a^{e}$ is even found to increase with $B$ in the BCS phase ($B<B_{c1}$). The reason for this increase is the enhancement in $F^{e}_{\text{max}}$ with $B$ is the presence of finite-energy pairs and the associated increase in the energy difference of these pairs with increasing $B$, as seen in Fig.~\ref{fig:pd}. At first glance, this magnetic field dependence of $K_a^{e}$  would seemingly suggest that the superconductor increases its superfluid weight with increasing $B$ and hence the superconductor becomes more stable in the magnetic field. However, this would be in sharp contradiction to the established notion that a magnetic field splits the spin Fermi surfaces and thus causes an energy cost in a spin-singlet superconductor. Notably, $K_a^{o}$, even if it is nearly three times smaller than $K_a^{e}$, exactly balances the increase in $K_a^{e}$, thus keeping the total $K_a$ unchanged for $B<B_{c1}$. Thus, appropriately including odd-frequency correlations is crucial for obtaining the correct magnetic field dependence of the superfluid weight in the presence of finite-energy pairs and to understand the stability of the superconductor. 

Moving on to higher magnetic fields, we find that the total anomalous contribution $K_a$ suffers a sudden reduction at the BCS to FF transition at $B_{c1}$ due to the sudden jump in the SC order parameter $\Delta^{Q}_0$, as seen in Fig.~\ref{fig:pd}. Thus in the FF phase the magnitude of $K_a^{o}$ becomes more comparable to the magnitude of $K_a^{e}$, but still the total contribution $K_a>0$. With further increase in $B$, $K_a$ further decreases and eventually reaches zero at the FF to normal transition at $B_{c2}$, as is expected. Finally, in Fig.~\ref{fig:meis}(b), we show the evolution of the superfluid weight $D_s$ and its individual components $K_a$ and $K_n$ with varying $B$. We find $K_a\approx K_n$ for all $B$. As a result, $D_s$ has the same $B$-dependence as $K_a$ and our above analysis in Fig.~\ref{fig:meis}(a) not just applies to $K_a$ but to the total superfluid weight.

The above analysis of the Meissner effect in the presence magnetic field shows the importance of SC correlations in experimental observables. In particular, we show that the odd-frequency correlations are essential to correctly describe the magnetic field dependence of the Meissner effect, especially when finite-energy pairs are present. 
It should here especially be emphasized that the common notion that odd-frequency correlations make the superconductor thermodynamically unstable due to a paramagnetic Meissner effect is thus not applicable as the total $K_a$ is never close to being negative. This thus provides the answer to the fourth and final question posed in Sec.~\ref{sec:Intro}. 

Finally we note that for the pair hopping model discussed in Sec.~\ref{sec:pairhopping}, the calculation of the Meissner effect becomes much more involved. The main reason is that the current operator in the pair hopping model contains higher order terms due to the modulating nature of the interaction \cite{Waardh17}. These higher order terms appear when the current operator is calculated using the continuity equation. The current-current correlation in Eq.~\eqref{eq:curcurgr} will thus involve many terms which substantially add to the complexity. Still, we believe the dominant contribution of the SC correlations to the superfluid weight is from the lowest order terms of the current operator and hence the Meissner effect in the pair hopping model should give qualitatively similar results as in Fig.~\ref{fig:meis}. 

\section{Conclusion and Discussion}\label{sec:Discussion}

In summary, in this work we first show that applying magnetic field to a conventional spin-singlet $s$-wave superconductor generates finite-energy Cooper pairs both in the low-field BCS phase and high-field finite-momentum FF phase. Our results thus illustrate that  finite-energy pairing exists even in the finite-momentum superconducting FF state. Furthermore, we find a direct connection between the odd-frequency SC correlations and finite-energy pairing by showing analytically that finite-energy pairs necessarily generate odd-frequency SC correlations. In contrast, even-frequency correlations originate primarily from zero-energy pairs for most frequencies, especially in the FF phase. We then study the interplay of finite-energy and finite-momentum Cooper pairs in a very different system, an unconventional $d$-wave superconductor with a spontaneous finite-momentum superconducting FF state driven by stripe formation even in the absence of any applied magnetic field. Here we find very similar relationships between odd-frequency SC correlations, finite-energy, and finite-momentum paring as in the magnetic-field driven FF state. These results establish that finite-energy and odd-frequency pairing are intimately linked and both prevalent in finite-momentum superconducting states. In particular, the formation of finite-momentum pairing does not remove either finite-energy or odd-frequency pairing.
Finally, we investigate the experimental consequences of the interplay between different variants of Cooper pairs by calculating the Meissner effect. Focusing only on the conventional $s$-wave superconductor under applied magnetic field, we show that odd-frequency correlations are necessary to correctly describe the magnetic field dependence of the superfluid weight or equivalently the Meissner effect.

Our finding of a close connection between  finite-energy pairs and odd-frequency SC correlations raises the question whether such a relation is also present in other systems where odd-frequency SC correlations are known to be present, but in the absence of finite-momentum superconductivity. Here we comment on such possibilities. Odd-frequency pairing is often discussed in the context of multiband superconductors \cite{Black-Schaffer13,Triola20}. In  multiband superconductors, odd-frequency correlations can also be connected to finite-energy pairing. For example, if we consider a two-band case with no intra-band pairing, such that the two individual bands can still be treated with individual normal state band dispersions $\xi_a$ and $\xi_b$, the inter-band odd-frequency pairing can be shown to be directly proportional to the difference of the energies of pairing electrons in the two bands, $\xi_a$-$\xi_b$ \cite{Black-Schaffer13}. Hence, also in multiband superconductors, odd-frequency correlations are directly related to the energy difference of the pairing electrons. A similar analogy can also be drawn for Ising superconductors. In such systems it has been shown that finite-energy Cooper pairs can be present in the ground state and odd-frequency correlations are then only present for Cooper pairs with finite-energy \cite{Tang21}.
A close connection between finite-energy pairs and odd-frequency SC correlations will likely also appear in a finite-momentum superconducting phase of both multiband and Ising superconductors. There exists for example already experimental evidence of a finite-momentum superconducting phase in the multiband superconductor FeSe \cite{Kasahara20} at high magnetic field. However, the additional presence of multiple bands and spin-orbit coupling may give additional interesting findings that can open new possible research directions. Another possible future prospect is to find a model for a finite-momentum pairing state in the absence of magnetic field in these systems, which to the best of our knowledge is still absent.

\begin{acknowledgments}

We thank M.~Granath for useful discussions. We gratefully acknowledge financial support from the Knut and Alice Wallenberg Foundation through the Wallenberg Academy Fellows program and the European Research Council (ERC) under the European Unions Horizon 2020 research and innovation programme (ERC-2017-StG-757553). The computations were enabled by resources provided by the Swedish National Infrastructure for Computing (SNIC) at the Uppsala Multidisciplinary Center for Advanced Computational Science (UPPMAX) partially funded by the Swedish Research Council through grant agreement no.~2018-05973.

\end{acknowledgments}

 \bibliographystyle{apsrev4-1}
\bibliography{Cuprates}

\end{document}